 \documentclass[prl,twocolumn,floatfix,superscriptaddress]{revtex4}
\usepackage{dcolumn,amsmath}
\usepackage{bm}

\newcommand{\eref}[1]{(\ref{#1})}

\newcommand{\tref}[1]{Table~\ref{#1}}

\begin{document}
\title{Configuration interaction calculation of hyperfine and P,T-odd constants
       on $^{207}$PbO excited states for the electron EDM experiments}.
\author{A.N.\ Petrov}\email{anpetrov@pnpi.spb.ru}
\author{A.V.\ Titov}
\author{T.A.\ Isaev}
\author{N.S.\ Mosyagin}
\affiliation
{Petersburg Nuclear Physics Institute, Gatchina,
             Leningrad district 188300, Russia}
\author{D.\ DeMille}
\affiliation
{Physics Department, Yale University, New Haven,
             Connecticut 06520, USA}


\begin{abstract}

 We report first configuration interaction calculations of hyperfine constants
 $A_{\parallel}$ and the effective electric field $W_d$ acting on the electric
 dipole moment of the electron, in two excited electronic states of
 $^{207}$PbO. A new combined scheme of correlation calculations is also 
 developed and first applied to studying the core properties.
 The obtained hyperfine constants, $A_{\parallel}= -3826$~MHz for the $a(1)$
 state and $A_{\parallel}= 4887$~MHz for the $B(1)$ state, are in very good
 agreement with the experimental data, $-4113$ {\rm MHz} and $5000 \pm 200$
 {\rm MHz}, respectively.
%
%
 We find $W_d = -\left( 6.1 { {^{+1.8}_{-0.6} } } \right) {\cdot}10^{24}\,{\rm
 Hz}/(e\cdot {\rm cm})$ for $a(1)$, and $W_d = -\left( { 8.0 \pm 1.6 } \right)
 {\cdot}10^{24}\,{\rm Hz}/(e\cdot {\rm cm})$ for $B(1)$.
%
\end{abstract}

\maketitle

 \paragraph*{Introduction.}
 The search for the electric dipole moment (EDM) of the electron, $d_e$,
 remains one of the most fundamental problems in physics.
 Up to now only upper limits were obtained for $|d_e|$; the latest constraint,
 $|d_e|<1.6\cdot10^{-27}$ e$\cdot$cm, was obtained in an experiment on atomic
 Tl \cite{Regan:02}.  The search for $d_e$ in heavy-atom polar molecules was
 initiated by theoretical investigations \cite{Sushkov:78,Gorshkov:79} on
 radicals with one unpaired electron. (The first experimental bound on $d_e$
 using such a system was recently obtained, using YbF \cite{Hudson:02}.)  It
 was since noted that experiments on the excited $a(1)$ \cite{DeMille:00} or
 $B(1)$ \cite{Egorov:01} states of PbO, which have two unpaired electrons, may
 make it possible to search for $d_e$ with sensitivity 2-4
 orders of magnitude higher than the current limit. An important feature of
 such experiments is that the knowledge of the effective electric field, $W_d$,
 seen by an unpaired electron is required for extracting $d_e$ from the
 measurements. $W_d$ can not be obtained in an experiment; rather, electronic
 structure calculations are required for its evaluation.  Recently, a
 semiempirical estimate of $W_d$ for the $a(1)$ state \cite{Kozlov:02} and {\it
 ab initio} relativistic coupled cluster (RCC) \cite{Kaldor:04ba} calculations
 \cite{Isaev:04} with the generalized relativistic effective core potential
 (GRECP) \cite{Titov:99, Titov:00a} of
 $W_d$ for both the $a(1)$ and $B(1)$ states
 were performed.  The nonvariational one-center
 restoration (NOCR) technique \cite{Titov:85Dis, Titov:96, Titov:96b, Titov:99}
 was employed for obtaining proper electronic densities close to nuclei after
 GRECP/RCC calculations of heavy-atom molecules.  The semiempirical estimate
 of Ref. \cite{Kozlov:02}, $|W_d|\ge 12\times 10^{24}$~Hz/($e\cdot$cm),
 is three times higher than the RCC result (see \tref{res1}).
 Calculations performed in \cite{Isaev:04} demonstrated the need for a higher
 level of accounting for correlation in the valence region for the excited
 $a(1)$ and $B(1)$ states. The main problem was that the Fock-space RCC with
 single and double excitations (RCC-SD) version used in \cite{Isaev:04} was not
 optimal for accounting for the nondynamic correlations (see \cite{Isaev:00}
 for details of RCC-SD calculations of the Pb atom), though the potential of
 the RCC method for electronic structure calculations is very high in prospect
 \cite{Kaldor:04ba}.  The estimated error bounds put the actual $W_d$ value
 between 75\% and 150\% of the value calculated by the GRECP/RCC/NOCR approach.

 The main goal of the present work is to develop a method of calculation which
 could provide substantially higher accuracy and reliability for $W_d$ in the
 excited states of PbO and could be efficiently applied to other heavy-atom
 molecules.  As an accuracy check for the calculation of the electronic
 structure near the Pb nucleus, the hyperfine constants ($A_{\parallel}$) are
 also calculated. For this purpose in the present work the {\sc sodci} code
 \cite{Buenker:99, Alekseyev:04a} for the spin-orbit direct configuration
 interaction (CI) calculation is employed, in which the relativistic
 (spin-orbit) scheme \cite{Titov:01} and new criteria \cite{Titov:01,
 Mosyagin:02} of the configuration selection proposed by us earlier are
 incorporated.  New codes for calculation of a one-electron density matrix with
 the CI wavefunction and restored four-component molecular spinors have been
 developed.  The 10-electron GRECP/SODCI/NOCR calculations are performed for
 $A_{\parallel}$ and $W_d$ in the $a(1)$ and $B(1)$ states of PbO with
 wavefunctions restored in the Pb core (see \cite{Mosyagin:98, Petrov:02,
 Isaev:04} for recent two-step calculations).

 The space of a SODCI calculation consists of a set of many-electron
 configurations obtained by selecting the singly- and doubly-excited
 configurations with respect to some multiconfigurational reference states when
 using some selection thresholds ($T$) \cite{Buenker:75, Titov:01}.  Because of
 limited computational resources, all the possible singly- and doubly-excited
 configurations (without selection) cannot be usually included to the SODCI
 calculation. The corrections accounting for zero threshold ($T{=}0$)
 \cite{Buenker:75} and higher-order excitations (``full-CI'') \cite{Bruna:80}
 allow one to estimate the contribution to the total energy of all the
 configurations omitted in the SODCI calculation (see below for more details).
 In the present work we first apply the
   $T{=}0$-, full-CI- and ``outercore correlations''-type corrections
 to calculating the properties which cannot be obtained
 from the energy curves (surfaces).
   The applicability of these corrections
   is justified and discussed in details in \cite{Titov:05c}.
 The developed method has allowed us to reduce the deviation of the calculated
 $A_{\parallel}$ from its experimental value for $a(1)$ in about five times,
 from 34\% to 7\%, as compared to the previous calculation \cite{Isaev:04}.
 That deviation for $B(1)$ is reduced from ($18\pm4$)\% to ($2\pm4$)\%, i.e.,
 it is now within the experimental uncertainty $\pm4$\%.  The SODCI results
 without applying the $T{=}0$, full-CI and outercore correlation corrections
 give the
 deviations 12\% for $a(1)$ and ($8\pm4$)\% for $B(1)$.  As a consequence,
 substantially more reliable values for $W_d$ are obtained.  Our present error
 estimate on $W_d$ is relatively more conservative than in \cite{Isaev:04}
 since
 $|W_d|$
 is increased from the RCC-SD datum on 50\% for $a(1)$ and decreased on 18\%
 for $B(1)$ (these notable changes are, however, within the error limits
 declared in \cite{Isaev:04}).
 The expression for $W_d$ is given by
 Refs.\,\cite{Kozlov:87, Kozlov:95, Isaev:04}

\begin{equation}
   W_d = \frac{1}{\Omega d_e}
   \langle \Psi_{X}|\sum_iH_d(i)|\Psi_{X} \rangle~,
\end{equation}
 where $ \Psi_{X} $ is the wavefunction for $a(1)$ or $B(1)$ state, and
 $\Omega= \langle\Psi_{X}|\bm{J}\cdot\bm{n}|\Psi_{X}\rangle = \pm1 $,
 $\bm{J}$ is the total electronic momentum, $\bm{n}$ is the unit vector
 along the molecular axis directed from Pb to O,

\begin{eqnarray}
    H_d=2d_e
    \left(\begin{array}{cc}
    0 & 0 \\
    0 & \bm{\sigma E} \\
    \end{array}\right),
\label{Wd}
\end{eqnarray}
 $\bm{E}$ is the inner molecular electric field, $\bm{\sigma}$ are the Pauli
 matrices. The hyperfine constant $A_{\parallel}$ is determined by the
 expression \cite{Dmitriev:92}
\begin{eqnarray}
   A_{\parallel}=\frac{1}{\Omega} \frac{\mu_{\rm Pb}}{I}
   \langle
   \Psi_{X}|\sum_i(\frac{\bm{\alpha}_i\times \bm{r}_i}{r_i^3})
_Z|\Psi_{X}
   \rangle~,
 \label{All}
\end{eqnarray}
 where $I$ and $\mu_{\rm Pb}$ is the spin and
 magnetic moment of $^{207}$Pb, $\bm{\alpha}_i$
 are the Dirac matrices for the $i$-th electron, and $\bm{r}_i$ is its
 radius-vector in the coordinate system centered on the Pb atom.

 \paragraph*{Methods and calculations.}
 The above properties were calculated in {two steps}.  At the {\it first step},
 a 22-electron GRECP for Pb \cite{Mosyagin:97} simulating interaction with the
 explicitly excluded $1s$ to $4f$ electrons is used.  In addition, the $5s^2
 5p_{1/2}^2 5p_{3/2}^4 5d_{3/2}^4 5d_{5/2}^6$ shells of lead and the $1s^2$
 shell of oxygen were frozen (see \cite{Titov:99} for details) and the residual
 ten electrons were treated explicitly in the subsequent molecular
 calculations. The basis set on Pb ($15s16p12d9f$)/[$5s7p4d2f$] as well as
 GRECP are the same as those used in paper \cite{Isaev:04}. The basis set was
 optimized for calculation of properties determined mainly by the electronic
 wave function at the Pb core. The description of the basis set generation
 procedure can be found in Refs.\ \cite{Isaev:00,Mosyagin:00}. The
 correlation-consistent ($10s5p2d1f$)/[$4s3p2d1f$] basis of Dunning listed in
 the {\sc molcas~4.1} library \cite{MOLCAS} was used for oxygen.

 The leading $\Lambda\Sigma$ coupling terms and configurations for the $a(1)$
 and $B(1)$ states are $^3\Sigma^+$\;$\sigma_1^2\sigma_2^2\sigma_3^2 \pi_1^3
 \pi_2^1$ and $^3\Pi_1$\;$\sigma_1^2\sigma_2^2\sigma_3^1 \pi_1^4 \pi_2^1$,
 respectively.  The molecular orbitals used in the CI calculations are obtained
 by the restricted active space self consistent field (RASSCF) method
 \cite{Olsen:88,MOLCAS} with the spin-averaged part of the GRECP (AGREP)
 \cite{Titov:99}, i.e.\ only scalar-relativistic effects are taken into account
 in the RASSCF calculation.  Because the $a(1)$ state is of primary interest,
 molecular orbitals were generated specifically for the lowest $^3\Sigma^+$
 state.
 This set of orbitals was used for the subsequent CI calculations of both the
 $a(1)$ and $B(1)$ states.  In the RASSCF method, orbitals are divided into
 three active subspaces: RAS1, with a restricted number of holes allowed; RAS2,
 where all possible occupations are included; and RAS3, with an upper limit on
 the number of electrons.  In this calculation, no more than two holes in RAS1
 and two electrons in RAS3 are allowed.  Using the $C_{2v}$ point group
 classification scheme, two A$_1$ orbitals in RAS1, six orbitals in RAS2, (two
 each in A$_1$, B$_1$, and B$_2$ irreps) and 50 (20 A$_1$, 6 A$_2$, 12 B$_1$,
 and 12 B$_2$) in RAS3 subspaces are included.

 Next the spin-orbit CI approach with the selected single- and
 double-excitations from some multiconfigurational reference states (``mains'')
 \cite{Buenker:74} is employed on the sets of different {$\Lambda$}S
 many-electron spin- and space-symmetry adapted basis functions (SAFs).  In the
 {\sc sodci} code, the double $C_{2v}$ group, $C_{2v}^*$, is used to account
 for the spin and space symmetry of the PbO molecule,  instead of the more
 restrictive symmetry group $C_{\infty v}^*$, which could in principle be
 employed. In the $C_{2v}^*$ classification scheme, the doubly-degenerate
 $a(1)$ and $B(1)$ states have the components in the $B_1^*$ and $B_2^*$
 irreducible representations (irreps).
 The operators $W_d$ and $A_{\parallel}$ have nonzero matrix elements only
 between the wavefunction components from $B_1^*$ and $B_2^*$ irreps.  So, one
 must know both these components of the $a(1)$ and $B(1)$ states to calculate
 $W_d$ and $A_{\parallel}$ when working in the $C_{2v}^*$ group.  The SAFs from
 the $^{2S{+}1}C_{2v}$-irreps (constructed on the basis of AGREP/RASSCF
 pseudoorbitals), singlet ($^1B_1$), triplets ($^3A_1$, $^3A_2$, $^3B_2$) and
 quintets ($^5A_1$, $^5A_2$, $^5B_1$, $^5B_2$), were included in calculations
 of the components belonging to the $B_1^*$ irrep and equivalent calculations
 were performed for those laying in the $B_2^*$ irrep.  (Alternatively, the
 components belonging to the $B_2^*$ irrep can be constructed by acting on
 those from the $B_1^*$ irrep by the operator of projection of the total
 electronic momentum on the molecular axis.)
 The reference space, $\{ \Phi^{(0)}_I \}_{I=1}^{N_0}$, consisted of
 $N_0=2517$~SAFs having the largest coefficients in the probing CI calculation.
 The single and double excitations from this reference space produce about
 $N_{T{=}0}{\approx}175\,000\,000$~SAFs, $\{ \Phi^{(1,2)}_I
 \}_{I=1}^{N_{T{=}0}}$.  The molecular Hamiltonian {\bf H} in our calculations
 is presented as

\begin{equation}
   {\bf H} = {\bf H}^0 + V\ ,
 \label{H0corrSO}
\end{equation}
 where ${\bf H}^0$ is an unperturbed spin-averaged
 Hamiltonian constructed to be diagonal in the given many-electron basis set:

\begin{equation}
    {\bf H}^0 \Phi^{(0,1,2)}_I\ =\
                      E^{(0,1,2)}_I \Phi^{(0,1,2)}_I\ ,
 \label{H0Psi}
\end{equation}
 and ${\bf V}$ is a perturbation that includes the
 two-electron operator describing
 correlations and a one-electron effective spin-orbit operator.
 Only the most important SAFs from $\{ \Phi^{(1,2)}_I\}$ set,
   which give contribution in the second-order perturbation theory {\it by
   energy} greater than some chosen threshold $T_i$:

\begin{equation}
     \frac{|<\Phi^{(1,2)}_J|{\bf V}|\Psi^{0}_{X} >|^2}
          {E^{(1,2)}_J - {\cal E}^{0}_{X}}\quad = \delta E_{XJ}
          \ge\quad T_i\ ,
 \label{n<Vcorr>^2}
\end{equation}
 were selected (with all SAFs from the $\{ \Phi^{(0)}_I \}$ set) for the
 subsequent CI calculation.  In the above equation, ${\cal E}^{0}_{X}$ and
 $\Psi^{0}_{X}$ are the eigenvalue and eigenfunction of ${\bf H}$ for $a(1)$ or
 $B(1)$ state in the subspace $\{ \Phi^{(0)}_I \}$.  About 120\,000, 500\,000,
 1\,100\,000, and 2\,000\,000 SAFs were selected (see \tref{res1}) for the
 thresholds (in $10^{-6}$\,a.u.) $T_1{=}0.1$, $T_2{=}0.01$, $T_3{=}0.0025$ and
 $T_4{=}0.001$, respectively.  (Selection on the CI {\it coefficients} obtained
 with the first-order perturbation theory \cite{Titov:01} was also used in our
 calculations of $A_{\parallel}$ and $W_d$ but the obtained results for the
 same number of selected SAFs were very close to those obtained using
 \eref{n<Vcorr>^2} and they are not presented in the paper.) The final
 wavefunction is written as

\begin{equation}
    \Psi^{T_i}_{X} = \sum_{n=0}^{2}\sum_{J} 
           C^{T_i(n)}_{XJ} \Phi^{(n)}_J,
\end{equation}
 where summation is performed over selected SAFs for a given threshold $T_i$
 and those from the $\{ \Phi^{(0)}_I \}$ set. With this wavefunction, energy
 and any property $W$ can be easily calculated:

\begin{equation}
     E^{T_n}_{X} =
         \langle\Psi^{T_n}_{X}|{\bf H}|\Psi^{T_n}_{X}\rangle,
\end{equation}
\begin{equation}
     W^{T_n}_{X} =
         \langle\Psi^{T_n}_{X}|{\bf W}|\Psi^{T_n}_{X}\rangle.
\end{equation}
 Then, the linear $T{=}0$ \cite{Buenker:75} and generalized Davidson (full-CI
 or FCI) \cite{Bruna:80} corrections are applied to the calculated properties.
 The $T{=}0$ correction extrapolates the results of calculations with two
 different thresholds $T_i$ and $T_j$ to a result of the calculation with all
 singly- and doubly-excited SAFs included. The full-CI correction approximates
 the effect of all possible excitations of higher level (for a given number of
 correlated electrons and fixed basis set). The $T{=}0$ correction for energy
 is given by

\begin{eqnarray}
 \label{T=0-cor}
      E^{T{=}0}_{X} = E^{T_n}_{X} +  \lambda
P^{T_n}_{X}\quad\Rightarrow \quad \\ 
\nonumber
      \lambda = - (E^{T_i}_{X}{-}E^{T_j}_{X})/(P^{T_i}_{X}{-}P^{T_j}_{X})\ ,
\end{eqnarray}
 where $n=i,j$ and
\begin{equation}
     P^{T_n}_{X} = \sum_{J:\ \delta E_{XJ} < T_n}
         \delta E_{XJ}\ .
\label{P}
\end{equation}
 The full-CI corrected energy is given by 

\begin{equation}
      E^{FCI}_{X} \approx E^{T{=}0}_{X} + (1 -
   |c^{\tt(0)}_I|^2) \cdot (E^{T{=}0}_{X} - {\cal E}^{0}_{X})\ ,
 \label{eFCI-cor}
\end{equation}
 where $|c^{\tt(0)}_I|^2$ are weights of the reference SAFs in
 $\Psi^{T_{\min}}_{X},~ T_{\min}=\min(T_i,T_j)$.  If all singly-excited SAFs,
 $\{ \Phi^{(1)}_I \}$, are included in calculation
    and the number of the reference SAFs is large enough,
 then similar expresions for $T{=}0$ and FCI corrections (in more details they
 will be discussed in \cite{Titov:05c}) for the {\it one-electron} property $W$
 can be written:

\begin{eqnarray}
\label{T=0-prop}
      W^{T{=}0}_{X} = W^{T_n}_{X} +  \lambda
P^{T_n}_{X}\quad\Rightarrow \quad \\
\nonumber
      \lambda = - (W^{T_i}_{X}{-}W^{T_j}_{X})/(P^{T_i}_{X}{-}P^{T_j}_{X})\ ,
\end{eqnarray}
 where $n=i,j$,

\begin{equation}
      W^{FCI}_{X} \approx W^{T{=}0}_{X} + (1 -
   |c^{\tt(0)}_I|^2) \cdot (W^{T{=}0}_{X} - {\cal W}^{0}_{X})\ ,
 \label{eFCI-corPro}
\end{equation}
 where ${\cal W}^{0}_{X}=\langle \Psi^{0}_{X}|{\bf W}|\Psi^{0}_{X}\rangle$. 
   Though equations \eref{T=0-prop} and \eref{eFCI-corPro} look similar to
   \eref{T=0-cor} and \eref{eFCI-cor}, they have essentially different physical
   sense.
  More details about the features of constructing the reference space and
  selection procedure are given in Refs.~\cite{Titov:01,Titov:05c}.

 Before calculating $A_{\parallel}$ and $W_d$, the shapes of the four-component
 molecular spinors are restored in the inner core region after the
 two-component GRECP calculation of the molecule. For this purpose the NOCR
 method \cite{Titov:85Dis} is applied at the {\it second step} of the
 calculation. (See \cite{Petrov:02} for the currently used formulation of the
 NOCR scheme and \cite{Isaev:04} for details of restoration of the
 four-component molecular spinors near the Pb nucleus).

 We designate the $5s,5p,5d$ orbitals of lead and $1s$ orbital of oxygen as
 the "outercore", and the $\sigma_1$, $\sigma_2$, $\sigma_3$, $\pi_1$, $\pi_2$
 orbitals of PbO (consisting mainly of $6s,6p$ orbitals of Pb and $2s,2p$
 orbitals of O) as valence. In the CI calculations we take into account only
 the correlations between valence electrons. At the final stage of our
 calculation, we estimate the contribution from correlations between valence
 and outercore electrons (including high-order correlations between outercore
 electrons) as difference in the results of the corresponding 30- and
 10-electron GRECP/RCC calculations. (See also \cite{Isaev:00} where this
 correction is applied to the Pb atom).  Such correlations are designated in
 the text and tables as ``outercore correlations''.

\paragraph*{Results and discussion.}
 Calculations were performed at two internuclear distances, $R=3.8$ a.u.\ (as
 in the RCC calculations of \cite{Isaev:04}), and $R=4.0$ a.u.  The latter
 point is closer to the equilibrium distances both for $a(1)$ ($R_e=4.06$
 a.u.~\cite{Martin:88}) and for $B(1)$ ($R_e=3.914$ a.u.~\cite{Huber:79}).
 However, in the RCC calculations \cite{Isaev:04}, $R=3.8$ a.u.\ was used
 because of a problem with convergence at $R=4.0$ a.u.  The calculated values
 with the one-center expansion of the molecular spinors in the Pb core on
 either $s$, $s;p$ or $s;p;d$ partial waves are collected in \tref{res1}.  The
 final data are obtained as the result of linear extrapolation to the
 experimental equilibrium distances.

 As was noted in \cite{Isaev:04} when comparing the RASSCF results with those
 obtained by the RCC method, the spin-orbit interaction changes $A_{\parallel}$
 and $W_d$ dramatically, so even the sign for $W_d$ is changed. One can see a
 similar picture from comparison of the RASSCF and CI data obtained in the
 present work.  The second point to notice is a significant difference of the
 results obtained at the internuclear distances $R=3.8$ and $R=4.0$ a.u.,
 especially, for the $W_d$ parameter. It is increasing for the $a(1)$ state and
 is decreasing for the $B(1)$ state, by 15\% and 30\%, respectively. The
 agreement with the experimental datum attained in the CI calculation of
 $A_{\parallel}$ for the $a(1)$ state at the point $R=4.0$ with the threshold
 $T_4{=}0.001$
 is 13\%.  After applying $T{=}0$ and FCI corrections and taking
 into account the outercore correlations the agreement is improved to the level
 of 9\%.  The calculated $A_{\parallel}$ value for the $B(1)$ state coincides
 with the measured datum within the experimental uncertainty of 4\%.  The
 results with the $T{=}0$ correction are close to those with the smallest
 threshold $T_4{=}0.001$.  To check the reliability of the linear $T{=}0$
 correction \cite{Buenker:75} for $A_{\parallel}$ and $W_d$ we have calculated
 it for three different pairs of thresholds: $T_1$ and $T_2$, $T_2$ and $T_3$,
 $T_3$ and $T_4$. For $A_{\parallel}$ all three pairs give the same result
 within 1\% of accuracy.  As to $W_d$, the result with the $T{=}0$ correction
 for the first pair differs from those for the other two pairs by 8\% for
 $a(1)$ and by 5\% for $B(1)$.  However, the last two pairs again give the same
 result within the accuracy of 1\% for both states.  So, we suggest, that the
 $T{=}0$ limit is determined for our main configurations with an accuracy of
 1\% for both considered properties.  Because the reference space is large
 enough, we have a small FCI correction.
 When taking into account outercore contributions at the point $R=4.0$ a.u.\ we
 used the results of the RCC calculation at the point $R=3.8$ a.u.
 This assumption seems reasonable for several reasons.  First, the core should
 relax less than the valence region when $R$ is changed.  In addition, because
 of the spatial separation between the core and valence electrons, core-valence
 correlation contributions should be more stable than valence-valence ones.
 Finally, since these contributions are relatively small, we expect errors due
 to this approximation not to be severe.

%
 We next discuss the uncertainty in the calculated $W_d$ values.  (We confine
 this detailed discussion to the $a(1)$ state; similar considerations are
 applied to $B(1)$.)  Since $W_d$ is sensitive to the wavefunction and its
 derivative at the Pb nucleus, it is natural to use the value of
 $A_{\parallel}$
 (which is also singular close to the Pb nucleus)
 as a benchmark for accuracy and stability. Thus, the
 10\% deviation of the calculated value of $A_{\parallel}$ from the
 experimental value represents an obvious lower bound for the accuracy of
 $W_d$.  It appears that $W_d$ is less computationally stable than
 $A_{\parallel}$, however, as may be noted from the variation in values at
 various stages of the calculation, shown in Table \ref{res1}.  Thus, this
 simple argument may underestimate the error in $W_d$.  Because of the good
 convergence of the CI calculation, it appears that the deviation in
 $A_{\parallel}$ can be due in part to sensitivity of the results to the value
 of $R$, and in part to incomplete accounting for outercore-valence
 correlations.
%
 Consideration of the former effect leads us to a possible error in $W_d$ of
 $\sim 10$\%.  The latter is more difficult to account for numerically, but
 (based on experience from similar calculations on HgF, PbF, YbF, and BaF
 \cite{Dmitriev:92,Titov:96b,Kozlov:97,Mosyagin:98}) we believe that the
 apparent insensitivity of $A_{\parallel}$ to outercore-valence correlations
 (compared to the case for $W_d$) is an artifact unlikely to persist in more
 advanced calculations.  Finally, we note that every improvement to the
 calculation (increasing correlation threshold; inclusion of core-valence
 correlation; increasing value of $R$ towards the equilibrium value) actually
 increases $|W_d|$.  Thus, the true value of $|W_d|$ is not likely to be much
 lower than the present calculated value.
 The linear extrapolation of the calculated $A_{\parallel}$ and $W_d$ to the
 experimental equilibrium distance for $a(1)$, $R_e=4.06$
 a.u.~\cite{Martin:88}, gives us $-3826$~MHz and $-6.1{\cdot}10^{24}{\rm
 Hz}/(e\cdot {\rm cm})$, respectively, and the deviation of the extrapolated
 $A_{\parallel}$ value from the experiment now is only 7\%.  Accounting for the
 above arguments, the estimated error bounds put the actual
 $W_d$ value between 90\% and 130\% of our final value.  We obtain $W_d$ for
 $B(1)$ by linear interpolation of the data obtained at $R=3.8$ and $4.0$ a.u.\
 to the experimental equilibrium distance, $R_e=3.914$ a.u.~\cite{Huber:79},
 yielding $W_d = -8.0{\cdot}10^{24}{\rm Hz}/(e\cdot {\rm cm})$, with an
 estimated error range of $\pm 20$\,\%.
 Similar interpolation for $A_{\parallel}$ on $B(1)$ gives $4887$~MHz, which is
 within the uncertainty of the current experimental value \cite{Kawall:04a},
 $5000 \pm 200 {\rm MHz}$.

\begin{table*}
\caption {
 Calculated  parameters $A_{\parallel}$ (in MHz) and $W_d$ (in $10^{24}{\rm
 Hz}/(e\cdot {\rm cm})$) for the $a(1)$ and $B(1)$ states of $^{207}$PbO at
 internuclear distances 3.8 and 4.0 a.u.  The experimental values of
 $A_{\parallel}$ are $-4113$\,MHz in $a(1)$ \cite{Hunter:02},
 and $5000 \pm 200$\,MHz in $B(1)$ \cite{Kawall:04a}.
}
\begin{ruledtabular}
\begin{tabular}{lrrrrrrrrrrrrr}
State &\multicolumn{6}{c}{$a(1)$\ \
$\sigma_1^2 \sigma_2^2 \sigma_3^2 \pi_1^3 \pi_2^1$ \ \
$^3\Sigma_1$}& &\multicolumn{6}{c}{$B(1)$\ \
$\sigma_1^2 \sigma_2^2 \sigma_3^1 \pi_1^4 \pi_2^1$\ \
$^3\Pi_1$}\\
Parameters &\multicolumn{3}{c}{$A_{\parallel}$}
&&\multicolumn{2}{c}{$W_d$}
& & \multicolumn{3}{c}{$A_{\parallel}$}&& \multicolumn{2}{c}{$W_d$}\\
\cline{2-4}\cline{6-7}\cline{9-11}\cline{13-14} Expansion & s&
s,p& s,p,d&& s,p& s,p,d& & s& s,p& s,p,d&& s,p&

s,p,d \\

\hline
\multicolumn{14}{c}{ }  \vspace{-2mm} \\
 \multicolumn{1}{c}{($T$ is in $10^{-6}$\,a.u.)} &
 \multicolumn{13}{c}{\bf Internuclear distance $R=3.8$ a.u. } \\
\multicolumn{14}{c}{ }  \vspace{-2mm} \\

 10e-RASSCF

& -894 & -1505 & -1503 && 0.73 & 0.70 & & & & &&0.0 &0.0 \\

\hline

 10e-RCC-SD \cite{Isaev:04}

&     &      & -2635  && -2.93& -3.05 & & & &3878&& -11.10 & -10.10 \\

 30e-RCC-SD \cite{Isaev:04}

&     &      & -2698  &&      & -4.10 & &    &     & 4081&& -9.10 &-9.70 \\

\it outercore (30e-RCC-SD - 10e-RCC-SD)

&     &      & \it -63&&     & \it -1.05& &    &  & \it 203&&  &
\it 0.40  \\

\hline 10e-CI (reference)$^a$
&-406 & -1877 & -1874  && -0.74 & -0.83 && 731 & 3785 & 3805 && -7.67 & -7.17 \\
10e-CI ($T{=}0.1$)
& -472& -2930 & -2926  && -2.12 & -2.21 && 393 & 4051 & 4074 && -9.85 & -9.40 \\
10e-CI ($T{=}0.01$) & -430& -3222 &  -3218 && -3.03 & -3.13 && 371
& 4320 & 4344 &&-10.16 & -9.72
\\
10e-CI ($T{=}0.0025$) & -412& -3304 & -3300  && -3.44 &  -3.54&&
359 & 4411 & 4436 &&-10.46 & -
10.02\\
10e-CI ($T{=}0.0012$) & -407&-3332  & -3328  && -3.58 &  -3.69&&
359 & 4449 & 4474 &&-10.62 & -
10.18\\

10e-CI + $T{=}0$
&     &       & -3387  &&       &  -4.01&&     &      & 4555 &&       & -10.52\\
\it 10e-CI + $T{=}0$ + FCI &     &       & \it -3446 &&    & \it
-4.13 &&     &  & \it 4582&&       & \it
-10.64 \\
\bf FINAL
&     &       &            &&       &           &&  & &         &&  &
\vspace{-1mm}       \\
(10e-CI + $T{=}0$ + FCI + outercore)
&     &       & \bf -3509 && & \bf -5.18  &&     & & \bf 4785 &&       & \bf -10.24 \\
\multicolumn{14}{c}{ }  \vspace{-2mm} \\
 \multicolumn{1}{c}{($T$ is in $10^{-6}$\,a.u.)} &
 \multicolumn{13}{c}{\bf Internuclear distance $R=4.0$ a.u. } \\
\multicolumn{14}{c}{ }  \vspace{-2mm} \\

 10e-RASSCF

& -770 & -1384 & -1383 && 1.05   & 1.00   & & & & &&0.0 &0.0 \\

\hline

10e-CI (reference)
& -459& -2026 & -2025  && -0.64 & -0.72 && 966 & 4127 & 4150 && -6.69 & -6.22 \\
10e-CI ($T{=}0.1$)
&-479 & -3125 & -3124  &&  -2.34& -2.44 && 525 & 4332 & 4357 && -7.79 & -7.35 \\
10e-CI ($T{=}0.01$)
&-449 & -3458 & -3458  && -3.50 & -3.61 && 495 & 4565 & 4590 && -7.38 & -6.94 \\
10e-CI ($T{=}0.0025$)
& -426& -3536 & -3536  && -3.97 & -4.08 && 481 & 4636 & 4662 && -7.45 & -7.02 \\
10e-CI ($T{=}0.001$)
& -422& -3571 & -3571  && -4.19 & -4.31 && 480 & 4666 & 4692 && -7.49 & -7.07 \\

10e-CI + $T{=}0$
&     &       & -3625  &&       & -4.65 &&     &      & 4739 &&       & -7.15 \\
\it 10e-CI + $T{=}0$ + FCI
&     &       & \it -3689  &&       & \it -4.81 &&  & &\it 4762 &&  &\it -7.18 \\
{\bf FINAL}
&     &       &            &&       &           &&  & &         &&  &
\vspace{-1mm}       \\
(10e-CI + $T{=}0$ + FCI + outercore)$^b$
&     &       & \bf -3752  &&       & \bf -5.86 &&  & &\bf 4965 &&  &\bf -6.78 \\

\end{tabular}
\end{ruledtabular}
\begin{flushleft}
 $^{\rm a}$ ``Reference'' means that the CI calculation was performed
 with the reference (main) SAFs only.\\
 $^{\rm b}$ It is assumed that the outercore contribution at the
 internuclear distance $R=4.0$ a.u.\ is approximately the same as is at the
 point $R=3.8$ a.u.
\end{flushleft}
 \label{res1}
\end{table*}


\paragraph*{Conclusion.}
 A new combined method for accurate calculating core properties
 in excited states of heavy-atom molecules is developed. It has allowed us to
 obtain accurate $A_{\parallel}$ and $W_d$ values for the $a(1)$ and $B(1)$
 states of PbO in spite of their complexity. The developed method and codes
 open an opportunity of reliable theoretical study of whole multitude of core
 properties in such computationally challenging objects as low-lying states of
 heavy-atom systems with valence $p$-electrons and dense spectra.

\paragraph*{Acknowledgments.}
 The present work is supported by U.S.\ CRDF grant RP2--2339--GA--02 and
 RFBR grant 03--03--32335.  A.P.\ is grateful to the Ministry of Education of
 the Russian Federation (grant PD02-1.3-236). T.I.\ thanks INTAS for grant YSF
 2001/2-164.  N.M.\ was supported by the grants of Russian Science Support
 Foundation and the governor of Leningrad district. D.D.\ acknowledges support
 from NSF grant PHY0244927 and the David and Lucile Packard Foundation.

\bibliographystyle{./bib/apsrev}


\bibliography{bib/JournAbbr,bib/Titov,bib/TitovLib,bib/Kaldor,bib/Isaev}

\end{document}